\documentclass[pre,twocolumn,superscriptaddress]{revtex4-2}

\usepackage{dcolumn}
\usepackage{amsmath}
\usepackage{amsfonts}
\usepackage{graphicx}
\usepackage{calrsfs}
\usepackage{bm}
\usepackage{enumitem}
\usepackage{hyperref}
\usepackage{mdwlist}
\usepackage{multirow}
\usepackage{soul}

\newcommand{\mat}{\bm}
\renewcommand{\vec}{\bm}
\newcommand\etal{\textit{et~al.}}

\newcommand{\Atest}{\mat{A}_{\text{test}}}
\newcommand{\Atrain}{\mat{A}_{\text{train}}}

\begin{document}
\title{Luck, skill, and depth of competition in games and social hierarchies}

\author{Maximilian Jerdee}
\affiliation{Department of Physics, University of Michigan, Ann Arbor, Michigan 48109, USA}

\author{M. E. J. Newman}
\affiliation{Department of Physics, University of Michigan, Ann Arbor, Michigan 48109, USA}
\affiliation{Center for the Study of Complex Systems, University of Michigan, Ann Arbor, Michigan 48109, USA}

\begin{abstract}
Patterns of wins and losses in pairwise contests, such as occur in sports and games, consumer research and paired comparison studies, and human and animal social hierarchies, are commonly analyzed using probabilistic models that allow one to quantify the strength of competitors or predict the outcome of future contests.  Here we generalize this approach to incorporate two additional features: an element of randomness or luck that leads to upset wins, and a ``depth of competition'' variable that measures the complexity of a game or hierarchy.  Fitting the resulting model to a large collection of data sets we estimate depth and luck in a range of games, sports, and social situations.  In general, we find that social competition tends to be ``deep,'' meaning it has a pronounced hierarchy with many distinct levels, but also that there is often a nonzero chance of an upset victory, meaning that dominance challenges can be won even by significant underdogs.  Competition in sports and games, by contrast, tends to be shallow and in most cases there is little evidence of upset wins, beyond those already implied by the shallowness of the hierarchy.
\end{abstract}

\maketitle

\section{Introduction}
\label{sec:intro}
One of the oldest and best-studied problems in data science is the ranking of a set of items, individuals, or teams based on the results of pairwise comparisons between them.  Such problems arise in sports, games, and other competitive human interactions, in paired comparison surveys in market research and consumer choice, in revealed-preference studies of human behavior, and in studies of social hierarchies in both humans and animals.  In each of these settings, one has a set of comparisons between pairs of items or competitors, with outcomes of the form ``A beats B'' or ``A is preferred to~B,'' and the goal is to determine a ranking of competitors from best to worst, allowing for the fact that the data may be sparse (there may be no data for many pairs) or contradictory (e.g., A~beats B beats C beats~A).  A group of chess players might play in a tournament, for example, and record wins and losses against each other.  Consumers might express preferences between pairs of competing products, either directly in a survey or implicitly through their purchases or other actions.  A flock of chickens might peck each other as a researcher records who pecked whom in order to establish the classic ``pecking order.''

\begin{figure}
\begin{center}
\includegraphics[width=8cm]{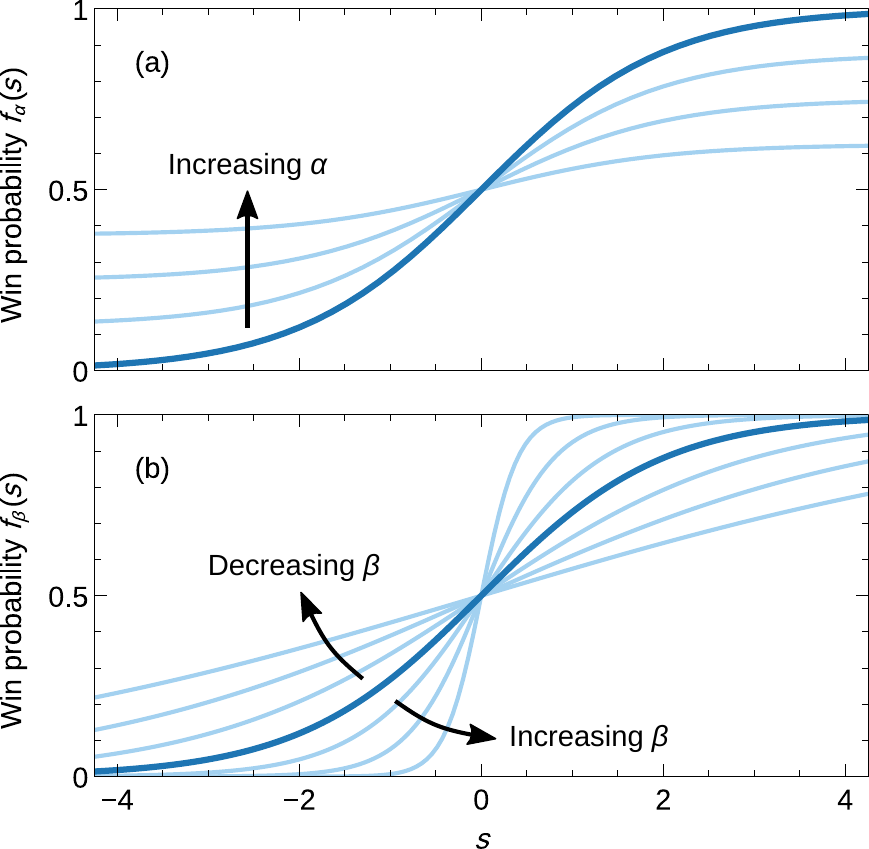}
\end{center}
\caption{Score functions~$f(s)$.  (a)~The bold curve represents the standard logistic function $f(s) = 1/(1+e^{-s})$ used in the Bradley-Terry model.  The remaining curves show the function~$f_\alpha$ of Eq.~\eqref{eq:fu} for increasing values of the luck parameter~$\alpha$.  (b)~The score function~$f_\beta$ of Eq.~\eqref{eq:fbeta} for different values of the depth of competition~$\beta$, both greater than~1 (steeper) and less than~1 (shallower).}
\label{fig:score}
\end{figure}

A large number of methods have been proposed for solving ranking problems of this kind---see Refs.~\cite{David88,Cattelan12,LM13} for reviews.  In this paper we consider one of the most common, which uses a statistical model for wins and losses and then fits that model to observed win/loss data.  In the most widely adopted version one considers a population of~$n$ competitors labeled by $i=1\ldots n$ and assigns to each a real score parameter~$s_i \in [-\infty,\infty]$.  Then the probability that $i$ beats $j$ in a single pairwise match or contest is assumed to be some function of the difference of their scores: $p_{ij} = f(s_i-s_j)$.  The function~$f(s)$ satisfies the following axioms:
\begin{enumerate*}
\item It is increasing in~$s$, since by definition a better competitor has a higher probability of winning than a worse one.
\item It tends to~1 as $s\to\infty$ and to 0 as $s\to-\infty$, meaning that an infinitely good player always wins and an infinitely poor one always loses.
\item It is antisymmetric about its mid-point at $s=0$, with the form
\begin{equation}
f(-s) = 1 - f(s),
\label{eq:antisym}
\end{equation}
because the probability of losing is, by definition, one minus the probability of winning.  As a corollary, this also implies that the probability~$f(0)$ of beating an evenly matched opponent is always~$\frac12$.
\end{enumerate*}
Subject to these constraints the function can still take a wide variety of forms, but the most popular choice by far is the logistic function $f(s) = 1/(1+e^{-s})$---shown as the bold curve in Fig.~\ref{fig:score}a---which gives
\begin{equation}
f(s_i-s_j) = {e^{s_i}\over e^{s_i}+e^{s_j}}.
\label{eq:f}
\end{equation}
The resulting model is known as the Bradley-Terry model, after R. Bradley and M. Terry who described it in 1952~\cite{BT52}, although it was (unbeknown to them) first introduced much earlier, by Zermelo in 1929~\cite{Zermelo29}.

Given the model, one can estimate the values of the score parameters~$s_i$ by a number of standard methods, including maximum likelihood estimation~\cite{Zermelo29,BT52,Ford57,Hunter04,Newman23b}, maximum a posteriori estimation~\cite{Whelan17}, or Bayesian methods~\cite{DS73,CD12}, then rank competitors from best to worst in order of their scores.  The fitted model can also be used to predict the outcome of future contests between any of the competitors, even if they have never directly competed in the past.

This approach is effective and widely used, but the standard Bradley-Terry model is a simplistic representation of the patterns of actual competition and omits many important elements found in real-world interactions.  Generalizations of the model have been proposed that incorporate some of these elements, such as the possibility of ties or draws between competitors~\cite{RK67,Davidson70}, multiway competition as in a horse race~\cite{Luce59,Plackett75}, or the ``home-field advantage'' of playing on your own turf~\cite{Agresti90}.  In this paper we consider a further extension of the model that incorporates two additional features of particular interest, which have received comparatively little previous attention: the element of luck inherent for instance in games of chance, and the notion of ``depth of competition,'' which captures the complexity of games or the number of distinct levels in a social hierarchy.  In the remainder of the paper we define and motivate this model and then describe a Bayesian approach for fitting it to data, which we use to infer the values of the luck and depth variables for a variety of real-world data sets drawn from different arenas of human and animal competition.  Our results suggest that social hierarchies are in general deeper and may have a larger element of luck to their dynamics than recreational games and sports, which tend to be shallower and show little evidence of a luck component. 

Software implementations of the various methods described in this paper are available at \verb|https://github.com/maxjerdee/pairwise-ranking|.

\section{The model}
\label{sec:model}
Suppose we observe $m$ matches between $n$ players.  The outcomes of the matches can be represented by an $n\times n$ matrix~$\mat{A}$ with element $A_{ij}$ equal to the number of times player~$i$ beats player~$j$.  Within the standard Bradley-Terry model the probability of a win is given by Eq.~\eqref{eq:f} and, assuming the matches to be statistically independent, the probability or likelihood of the complete set of observed outcomes is
\begin{align}
P(\mat{A}|\vec{s}) = \prod_{ij} f(s_i-s_j)^{A_{ij}}
         = \prod_{ij} \biggl( {e^{s_i}\over e^{s_i}+e^{s_j}} \biggr)^{A_{ij}},
\label{eq:PAGsf}
\end{align}
where $\vec{s}$ is the vector with elements~$s_i$.  (We assume that the structure of the tournament---who plays whom---is determined separately, so that~\eqref{eq:PAGsf} is a distribution over the directions of the wins and losses only and not over which pairs of players competed.)

The scores are traditionally estimated by the method of maximum likelihood, maximizing~\eqref{eq:PAGsf} with respect to all~$s_i$ simultaneously to give estimates
\begin{equation}
\hat{\vec{s}} = \text{argmax}_{\vec{s}} P(\mat{A}|\vec{s}).
\label{eq:sML}
\end{equation}
These maximum likelihood estimates (MLEs) can then be sorted in order to give a ranking of the competitors, or simply reported as measures of strength in their own right.  The widely used Elo ranking system for chess players, for example, is essentially a version of this approach, but extended to allow for dynamic updates as new matches are added to the data set.

The maximum likelihood approach unfortunately has some drawbacks.  For one, the likelihood is invariant under a uniform additive shift of all scores~$s_i$ and hence the scores are not strictly identifiable, though this issue can easily be fixed by normalization.  A more serious problem is that the likelihood maximum does not exist at all unless the network of interactions---the directed network with adjacency matrix~$\mat{A}$---is strongly connected (meaning there is a directed chain of victories from any player to any other), and the maximum likelihood estimation procedure fails, with the divergence of some or all of the scores, unless this relatively stringent condition is met.

This issue can be addressed by introducing a prior on the scores and adopting a Bayesian perspective.  A~variety of potential priors for this purpose have been systematically examined by Whelan~\cite{Whelan17}, who, after careful consideration, recommends a Gaussian prior with mean zero.  The variance is arbitrary---it merely sets the scale on which the score~$s$ is measured---but for subsequent convenience we here choose a variance of~$\frac12$ so that the prior on $\vec{s}$ takes the form
\begin{equation}
P(\vec{s}) = \prod_{i=1}^n \frac{1}{\sqrt{\pi}} \,e^{-s_i^2}.
\label{eq:sprior}
\end{equation}
An alternative prior, also recommended by Whelan, is the logistic distribution
\begin{align}
P_L(\vec{s}) = \prod_{i = 1}^n \frac{1}{(1 + e^{s_i})(1+e^{-s_i})}.
\label{eq:spriorlogistic}
\end{align}
In practice the Gaussian and logistic distributions are similar in shape and the choice of one or the other does not make a great deal of difference.  The logistic distribution is perhaps the less natural of the two and we primarily use the Gaussian distribution in this paper, but the logistic distribution does have the advantage of leading to faster numerical algorithms and we have used it in previous work for this reason~\cite{Newman22,Newman23b}.  We also include it in the basket of models that we compare in Section~\ref{sec:outcomes}.

Once we have defined a prior on the scores we can calculate a maximum a posteriori (MAP) estimate of their values as
\begin{align}
\hat{\vec{s}} = \text{argmax}_{\vec{s}} P(\vec{s}|\mat{A}) = \text{argmax}_{\vec{s}} P(\mat{A}|\vec{s})P(\vec{s}).
\label{eq:sMAP}
\end{align}
The MAP estimate always exists regardless of whether the interaction network is strongly connected, and using a prior also eliminates the invariance of the probability under an additive shift and hence the need for normalization.  As an alternative to computing a MAP estimate we can also simply return the full posterior distribution $P(\vec{s}|\mat{A})$, which gives us complete information on the expected values and uncertainty of the scores given the observed data.

\section{Extensions of the model}
\label{sec:extensions}
In this section we define generalizations of the Bradley-Terry model that extend the score function~$f$ in two useful ways, while keeping other aspects of the model fixed, including the normal prior.  The specific generalizations we consider involve dilation or contraction of the score function in the vertical and horizontal directions.  Vertical variation controls the element of luck that allows a weak player to sometimes beat a strong one; horizontal variation controls the ``depth of competition,'' a measure of the complexity of a game or contest.

\subsection{Upset wins and luck}
\label{sec:luck}
The first generalization of the Bradley-Terry model that we consider is one where the function~$f$ is contracted in the vertical direction, as shown in Fig.~\ref{fig:score}a.  We parametrize this function in the form
\begin{equation}
f_\alpha(s) = \tfrac12 \alpha + (1-\alpha) {1\over1+e^{-s}},
\label{eq:fu}
\end{equation}
with $\alpha\in[0,1]$.  In the traditional Bradley-Terry model~$f(s)$ tends to 0 and~1 as $s\to\pm\infty$, as discussed in the introduction, but in the modified model with $\alpha>0$ this is no longer the case.  One can think of the parameter~$\alpha$ as controlling the probability of an ``upset win'' in which an infinitely good player loses or an infinitely bad player wins.  (The probabilities of these two events must be the same because of the antisymmetry condition, Eq.~\eqref{eq:antisym}.)

For some games or competitions it is reasonable that $f(s)$ tends to zero and one at the limits.  In a game like chess that has no element of randomness, an infinitely good player may indeed win every time.  In a game of pure luck like roulette, on the other hand, both players have equal probability~$\frac12$ of winning, regardless of skill.  These two cases correspond to the extreme values $\alpha=0$ and $\alpha=1$ respectively in Eq.~\eqref{eq:fu}.  Values in between represent games that combine both luck and skill, like poker or backgammon, with the precise value of~$\alpha$ representing the proportion of luck.  For this reason we refer to $\alpha$ as the luck parameter, or simply the ``luck.''

(One could also consider the chance of the weaker player winning in the standard Bradley-Terry model to be an example of luck or an upset win, but that is not how we use these words here.  In the present context the ``luck''~$\alpha$ describes the probability of winning the game even if one's opponent is infinitely good, which is zero in the standard model but nonzero in the model of Eq.~\eqref{eq:fu} with $\alpha>0$.)

Another way to think about $\alpha$ is to imagine a game as a mixture of a luck portion and a skill portion.  With probability~$\alpha$ the players play a game of pure chance in which the winner is chosen at random, for instance by the toss of a coin.  Alternatively, with probability $1-\alpha$, they play a game of skill, such as chess, and the winner is chosen with the standard Bradley-Terry probability.  The overall probability of winning is then given by Eq.~\eqref{eq:fu} and the parameter~$\alpha$ represents the fraction of time the game is decided by pure luck.  By fitting~\eqref{eq:fu} to observed win-loss data we can learn the luck inherent in a competition or hierarchy.  We do this for a variety of data sets in Section~\ref{sec:results}.

\subsection{Depth of competition}
\label{sec:depth}
The second generalization we consider is one where the function~$f$ is dilated or contracted in the horizontal direction, as shown in Fig.~\ref{fig:score}b, by a uniform factor~$\beta>0$ thus:
\begin{equation}
f_\beta(s) = \frac{1}{1+e^{-\beta s}}.
\label{eq:fbeta}
\end{equation}
The slope of this function at $s=0$ is given by
\begin{equation}
f_\beta'(0) = \biggl[ {\beta e^{-\beta s}\over(1+e^{-\beta s})^2} \biggr]_{s=0}
  = \tfrac14 \beta,
\end{equation}
so $\beta$ is simply proportional to the slope.  A more functional way of thinking about $\beta$ is in terms of the probability that the stronger of a typical pair of competitors will win.  With a normal prior on $s$ of variance~$\frac12$ as described in Section~\ref{sec:model}, the difference $s_i-s_j$ between the scores of a randomly chosen pair of competitors will be a priori normally distributed with variance~1, meaning the scores will be separated by an average (root-mean-square) distance of~1.  Consider two players separated by this average distance.  If $\beta$ is small, making $f_\beta$ a relatively flat function (the shallowest curve in Fig.~\ref{fig:score}b), the probability $p_{ij}$ of the stronger player winning will be close to~$\frac12$ and there is a substantial chance that the weaker player will win.  Conversely, if $\beta$ is large then $p_{ij}$ will be close to~1 (the steepest curve in Fig.~\ref{fig:score}b) and the stronger player is very likely to prevail.

Thus one way to understand the parameter~$\beta$ is as a measure of the imbalance in strength or skill between the average pair of players.  When $\beta$ is large the contestants in the average game are very unevenly matched.  As we will shortly see, this is a common situation in social hierarchies, but not in sports and games, perhaps because contests between unevenly matched opponents are less rewarding both for spectators and for the competitors themselves.

Another way to think about $\beta$ is in terms of the number of levels of skill or strength in a competition.  Suppose we define one ``level'' as the distance $\Delta s = s_i-s_j$ between scores such that $i$ beats $j$ with a certain probability~$q$.  For a win probability of the form of Eq.~\eqref{eq:fbeta} we have $q = 1/(1+e^{-\beta\Delta s})$ and hence
\begin{equation}
\Delta s = {1\over \beta} \log {q\over1-q}. \label{eq:Deltas}
\end{equation}
Considering again the typical pair of players a distance~1 apart, the number of levels between them is
\begin{equation}
{1\over\Delta s} = {\beta\over\log[q/(1-q)]}.
\end{equation}
Thus the number of levels is simply proportional to~$\beta$.  Let us choose the probability~$q$ such that the constant of proportionality is~1, meaning $\log[q/(1-q)] = 1$ or
\begin{equation}
q = {1\over1+e^{-1}} = 0.731\ldots \label{eq:levelDefinition}
\end{equation}
With this definition, a ``level'' is the skill difference~$\Delta s$ between two players such that the better one wins 73\% of the time and our parameter~$\beta$ is simply equal to the number of such levels between the average pair of players.

In this interpretation, $\beta$~can be thought of as a measure of the complexity or depth of a game or competition.  A~``deep'' game, in this sense, is one that can be played at many levels, with players at each level markedly better than those at the level below.  Chess, which is played at a wide range of skill levels from beginner to grandmaster, might be an example.

This concept of depth has a long history.  For example, in an article in the trade publication \textit{Inside Backgammon} in 1980~\cite{Robertie80}, world backgammon champion William Robertie defined a ``skill differential'' as the strength difference between two players that results in the better one winning 70 to 75\% of the time---precisely our definition of a ``level''---and the ``complexity number'' of a sport or game as the number of such skill differentials that separate the best player from the worst.  Cauwet~\etal~\cite{Cauwet15} have defined a similar but more formal measure of game depth that they call ``playing-level complexity.''  There has also been discussion in the animal behavior literature of the ``steepness'' of animal dominance hierarchies~\cite{DSV06}, which appears to correspond to roughly the same idea.

One should be careful about the details.  Robertie and Cauwet~\etal\ both define their measures in terms of the skill range between the best and worst players, but this could be problematic in that the range will depend on the particular sample of players one has and will tend to increase as the sample size gets larger, which seems undesirable.  Our definition avoids this by considering not the best and worst players in a competition but the average pair of players, which gives a depth measure that is asymptotically independent of sample size.

Even when defined in this way, however, the number of levels is not solely about the intrinsic complexity of the game, but does also depend on who is competing.  For example, if a certain competition is restricted to contestants who all fall in a narrow skill range, then $\beta$ will be small even for a complex game.  In a world-class chess tournament, for instance, where every player is an international master or better, the number of levels of play will be relatively small even though chess as a whole has many levels.  Thus empirical values of~$\beta$ combine aspects of the complexity of the game with aspects of the competing population.

For this reason we avoid terms such as ``complexity number'' and ``depth of game'' that imply a focus on the game alone and refer to $\beta$ instead as the ``depth of competition,'' which we feel better reflects its meaning.

\subsection{Combined model}
\label{sec:combined}
Combining both the luck and depth of competition variables into a single model gives us the score function
\begin{equation}
f_{\alpha\beta}(s) = \tfrac12 \alpha + (1-\alpha) {1\over1+e^{-\beta s}}.
\label{eq:fab}
\end{equation}
In Section~\ref{sec:results} we fit this form to observed data from a range of different areas of study in order to infer the values of $\alpha$ and~$\beta$.  In the process one can also infer the scores~$s_i$, which can be used to rank the participants or predict the outcome of unobserved contests, and we explore this angle in Section~\ref{sec:outcomes}.  In this section, however, our primary focus is on $\alpha$ and $\beta$ and on understanding the varying levels of luck and depth in different kinds of competition.

\begin{table*}
\begin{center}
\begin{tabular}{l|lrrrll}
& Data set & $\hat{\beta}$ & $n$ & $m$ & Description & Ref. \\
\hline
\hline
\multirow{7}{*}{\rotatebox[origin=c]{90}{\hspace{10pt}Sports/games}}
& Scrabble       & 0.68 & 587 & 23477 & \textit{Scrabble} tournament matches 2004--2008 & \cite{Scrabble}\\
& Basketball     & 1.01 & 240 & 10002 & National Basketball Association games 2015--2022 & \cite{Basketball}\\ 
& Chess          & 1.17 & 917 & 7007 & Online chess games on lichess.com in 2016 & \cite{Chess}\\
& Tennis         & 1.44 & 1272 & 29397 & Association of Tennis Professionals matches 2010--2019 & \cite{Tennis}\\
& Soccer         & 1.73 & 1976 & 7208 & Men's international association football matches 2010--2019 & \cite{Soccer}\\
& Video games    & 1.77 & 125 & 1951 & \textit{Super Smash Bros Melee} tournament matches in 2022 & \cite{SSBM}\\
\hline
\multirow{3}{*}{\rotatebox[origin=c]{90}{Human}}
& Friends        & 3.54 & 774 & 2799 & High-school friend nominations & \cite{AddHealth}\\
& CS departments & 4.25 & 205 & 4388 & Doctoral graduates of one department hired as faculty in another & \cite{CAL15}\\
& Business depts. & 4.36 & 112 & 7856 & Doctoral graduates of one department hired as faculty in another & \cite{CAL15}\\
\hline
\multirow{6}{*}{\rotatebox[origin=c]{90}{Animal}}
& Vervet monkeys & 6.01 & 41 & 2930 & Dominance interactions among a group of wild vervet monkeys  & \cite{VBHB20}\\
& Dogs           & 8.74 & 27 & 1143 & Aggressive behaviors in a group of domestic dogs & \cite{SCCNM19}\\
& Baboons        & 13.19 & 53 & 4464 & Dominance interactions among a group of captive baboons & \cite{FMTAA15}\\
& Sparrows       & 22.92 & 26 & 1238 & Attacks and avoidances among sparrows in captivity & \cite{Watt86}\\
& Mice           & 26.48 & 30 & 1230 & Dominance interactions among mice in captivity & \cite{WFC16}\\
& Hyenas         & 100.58 & 29 & 1913 & Dominance interactions among hyenas in captivity & \cite{SH19}\\
\end{tabular}
\end{center}
\caption{Data sets analyzed in Section~\ref{sec:results}, in order of increasing depth of competition~$\beta$.  Here $n$ is the number of participants and $m$ is the number of matches/interactions.  Further information on the data sets is given in Appendix~\ref{app:datasets}.}
\label{tab:data}
\end{table*}

To perform the fit we consider again a data set represented by its adjacency matrix~$\mat{A}$ and write the data likelihood in the form of Eq.~\eqref{eq:PAGsf}:
\begin{equation}
P(\mat{A}|\vec{s},\alpha,\beta) = \prod_{ij} f_{\alpha\beta}(s_i-s_j)^{A_{ij}}.
\label{eq:likelihood}
\end{equation}
The scores~$\vec{s}$ are assumed to have the Gaussian prior of Eq.~\eqref{eq:sprior}, and we assume a uniform (least informative) prior on $\alpha$, which means $P(\alpha)=1$.  We cannot use a uniform prior on~$\beta$, since it has infinite support, so instead we use a prior that is approximately uniform over ``reasonable'' values of~$\beta$ and decays in some slow but integrable manner outside this range.  A suitable choice in the present case is (the positive half of) a Cauchy distribution centered at zero:
\begin{equation}
P(\beta) = {2w/\pi\over\beta^2+w^2}, \label{eq:Pbeta}
\end{equation}
where $w$ controls the scale on which the function decays.  In this paper we use $w=4$, which roughly corresponds to the range of variation in $\beta$ that we see in real-world data sets, and has the convenient property of giving a uniform prior on the angle of~$f_\beta(s)$ at the origin.

It is worth mentioning that the choice of prior on~$\beta$ \emph{does} have an effect on the results in some cases.  When data sets are large and dense, priors tend to have relatively little impact because the posterior distribution is narrowly peaked around the same set of values no matter what choice we make.  But some of the data sets we study here are quite sparse and for these the results can vary with the choice of prior.  Our qualitative conclusions remain the same in all cases, but it is worth bearing in mind that the quantitative details can change.

Combining the likelihood and priors, we now have
\begin{equation}
P(\vec{s},\alpha,\beta|\mat{A}) = P(\mat{A}|\vec{s},\alpha,\beta) {P(\alpha) P(\beta) P(\vec{s})\over P(\mat{A})}.
\label{eq:marginal}
\end{equation}
The prior on $\mat{A}$ is unknown but constant, so it can be ignored.  We now draw from the distribution $P(\vec{s},\alpha,\beta|\mat{A})$ to obtain a representative sample of values $\vec{s}, \alpha, \beta$.  In our calculations we generate the samples using the Hamiltonian Monte Carlo method~\cite{Neal11} as implemented in the probabilistic programming language Stan~\cite{Betancourt17}, which is ideal for sampling from continuous parameter spaces such as this.  A few thousand samples are typically sufficient to get a good picture of the distribution of $\alpha$ and~$\beta$.

\subsection{Minimum violations ranking}
One special case of our model worth mentioning is the limit $\beta\to\infty$ for fixed $\alpha>0$.  In this limit the function~$f_{\alpha\beta}(s)$ becomes a step function with value
\begin{equation}
f_{\alpha,\infty}(s) = \begin{cases}
  \tfrac12 \alpha   & \quad\mbox{if $s<0$,} \\
  \tfrac12          & \quad\mbox{if $s=0$,} \\
  1-\tfrac12 \alpha & \quad\mbox{if $s>0$.}
\end{cases} \label{eq:fainfty}
\end{equation}
For this choice the data likelihood becomes
\begin{equation}
P(\mat{A}|\mat{s},\alpha,\beta) = \bigl(\tfrac12\alpha\bigr)^v \bigl(1-\tfrac12\alpha\bigr)^{m-v},
\end{equation}
where $m$ is the total number of games/interactions/com\-parisons and $v$ is the number of ``violations,'' meaning games where the weaker player won.  Then the log-likelihood is
\begin{align}
\log P(\mat{A}|\mat{s},\alpha,\beta) &= -v \log {1-\frac12\alpha\over\frac12\alpha} + m \log\bigl(1-\tfrac12\alpha\bigr)
  \nonumber\\
  &= -Av - B,
\label{eq:MVR}
\end{align}
where $A$ and $B$ are positive constants.  This log-likelihood is maximized when the number of violations $v$ is minimized, which leads to the so-called \textit{minimum violations ranking}, the ranking such that the minimum number of games are won by the weaker player.  Thus the minimum violations ranking can be thought of as the limit of our model in the special case where $\beta\to\infty$.

\begin{figure*}
\centering
\includegraphics[width=\linewidth]{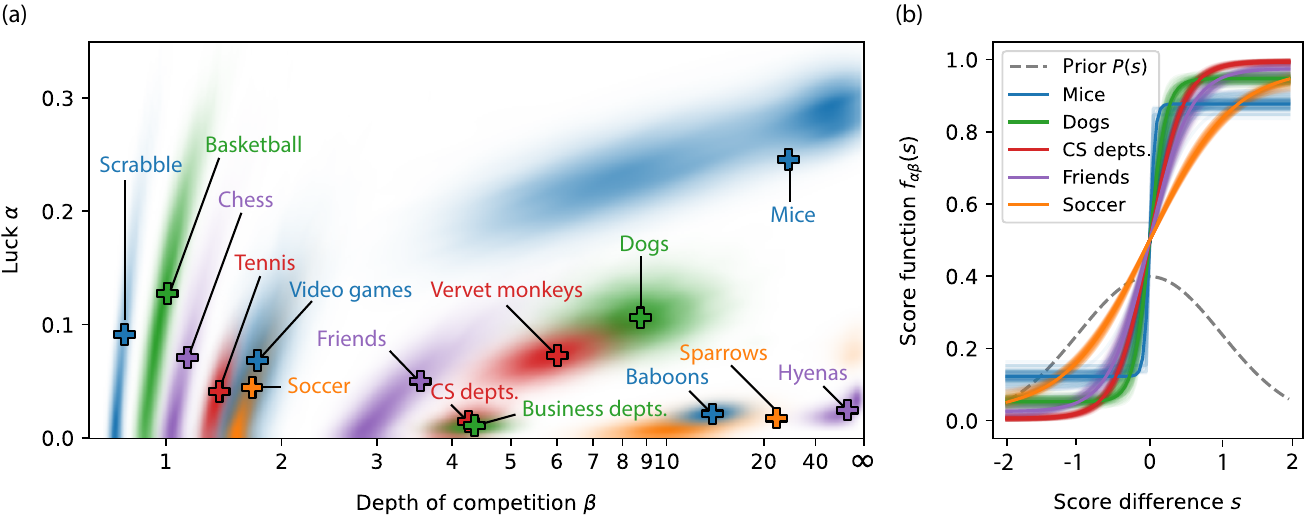}
\caption{(a)~Each cloud represents the posterior distribution~$P(\alpha,\beta|\mat{A})$ of the luck and depth parameters for a single data set, calculated from the Monte Carlo sampled values of $\alpha$ and $\beta$ using a Gaussian kernel density estimate.  The + signs indicate the expected values $\hat{\alpha},\hat{\beta}$ of the parameters for each data set.  (b)~Fitted functions~$f_{\alpha\beta}(s)$ for a selection of the data sets.  The bold curve in each case corresponds to the expected values $\hat{\alpha},\hat{\beta}$, while the other surrounding curves are for a selection of values sampled from the posterior distribution, to give an idea of the variation around the average.}
\label{fig:results}
\end{figure*}

\section{Results}
\label{sec:results}
We have applied these methods to a range of data sets representing competition in sports and games as well as social hierarchies in both humans and animals.  The data sets we study are listed in Table~\ref{tab:data}.

Figure~\ref{fig:results}a summarizes our results for the posterior probability density of the luck and depth parameters.  The axes of the figure indicate the values of $\alpha$ and~$\beta$ and each cloud is a Gaussian kernel density estimate of $P(\alpha,\beta|\mat{A})$ computed from the sampled values of $\alpha$ and~$\beta$ for a single data set.  The + signs in the figure represent the mean values of $\alpha$ and $\beta$ for each data set computed directly by averaging the samples.

The figure reveals some interesting trends.  Note first that all of the sports and games---chess, basketball, video games, etc.---appear on the left-hand side of the plot in the region of low depth of competition, while all the social hierarchies are on the right with higher depth.  We conjecture that the low depth of the sports and games is a result of a preference for matches to be between roughly evenly matched opponents, as discussed in Section~\ref{sec:depth}.  For a game to be entertaining to play or watch the outcome of matches should not be too predictable, but in a sport or league with high depth the average pairing is very uneven, with the stronger player very likely to win.  Low depth of competition ensures that matches are unpredictable and hence entertaining.  In games such as chess, which have high intrinsic depth, the depth can be reduced by restricting tournaments to players in a narrow skill range, such as world-class players, and this is commonly done in many sports and games.  We explore this interpretation further in Appendix~\ref{app:depth-predictability}.

There are no such considerations at play in social hierarchies.  Such hierarchies are not, by and large, spectator sports, and there is nothing to stop them having high depth of competition.  The results in Fig.~\ref{fig:results}a indicate that in general they do, though the animal hierarchies are deeper than the human ones.  A high depth in this context indicates a hierarchy in which the order of dominance between the typical pair of competitors is clear.  This accords with the conventional wisdom concerning hierarchies of both humans and animals, where it appears that participants are in general clear about the rank ordering.

Another distinction that emerges from Fig.~\ref{fig:results}a is that the results for sports and games generally do not give strong support to a nonzero luck parameter.  The expected values, indicated by the + signs, are nonzero in most cases, but the clouds representing the posterior distributions give significant weight to points close to the $\alpha=0$ line, indicating that we cannot rule out the possibility that $\alpha=0$ in these competitions.  For many of the social hierarchies, on the other hand, there is strong evidence for a nonzero amount of luck, with the posterior distribution having most of its weight well away from $\alpha=0$.

In part this observation is constrained by the data we have available.  It is difficult to distinguish the value of $\alpha$ in a competition with low depth because most matches are fairly evenly balanced---neither player is strongly favored to win.  We can also achieve the same outcome by making the luck parameter~$\alpha$ large, so that high luck and low depth both give good fits to the data and hence are confounded in the results.  This is reflected by the tall shapes of the posterior clouds on the left of Fig.~\ref{fig:results}a, indicating a high uncertainty about the value of~$\alpha$.  In the high-depth region on the right of the figure it is much easier to discern the value of~$\alpha$, and in this region there are many data sets for which we can be quite certain that $\alpha$ is nonzero.  This finding of nonzero $\alpha$ also accords with our intuition about social hierarchies.  There would be little point in having any competition at all within a social hierarchy if the outcomes of all contests were foregone.  If all participants knew that every competitive interaction was going to end with the higher-ranked individual winning and the lower-ranked one backing down, then there would be no reason to compete.  It is only because there is a significant chance of a win that competition occurs at all.

An interesting counter-example to this observation comes from the two faculty hierarchies, which represent hiring practices at US universities and colleges.  The interactions in this data set indicate when one university hires a faculty candidate who received their doctoral training at another university, which is considered a win for the university where the candidate trained.  The high depth of competition and low luck parameter for these data sets indicates that there is a pronounced hierarchy of hiring with a clear pecking order and that the pecking order is rarely violated.  Lower-ranked universities hire the graduates of higher-ranked ones, but the reverse rarely happens.

Figure~\ref{fig:results}b shows a selection of the fitted functions~$f_{\alpha\beta}(s)$ for five of the data sets.  For each data set we show in bold the curve for the expected values~$\hat\alpha,\hat\beta$ along with ten other curves for values of $\alpha,\beta$ sampled from the posterior distribution, to give an indication of the amount of variation around the average.  We see for example that the curve for the soccer data set has a shallow slope (low depth of competition) but is close to zero and one at the limits (low luck).  The curve for the mice data set, by contrast, is steep (high depth) but clearly has limits well away from zero and one (nonzero luck).

\section{Predicting wins and losses}
\label{sec:outcomes}
In addition to allowing us to infer the luck and depth parameters and rank competitors, our model can also be used to predict the outcomes of unobserved matches.  If we fit the model to data from a group of competitors, we can use the fitted model to predict the winner of a new contest between two of those same competitors.  The ability to accurately perform such predictions can form the basis for consumer product recommendations and marketing, algorithms for guiding competitive strategies in sports and games, and the setting of odds for betting, among other things.

We can test the performance of our model in this prediction task using a cross-validation approach.  For any data set $\mat{A}$ we randomly remove or ``hold out'' a small portion of the matches or interactions and then fit the model to the remaining ``training'' data set.  Then we use the fitted model to predict the outcome of the held-out matches and compare the results with the actual outcomes of those same matches.

The simplest version of this calculation involves fitting our model to the training data by making point estimates of the parameters and scores.  We first estimate the expected posterior values $\hat{\alpha},\hat{\beta}$ of the parameters given the training data.  Then, given these parameter values, we maximize the posterior probability as a function of~$\vec{s}$ to obtain MAP estimates~$\hat{\vec{s}}$ of the scores.  Finally, we use the combined parameter values and scores to calculate the probability~$\hat{p}_{ij} = f_{\hat\alpha\hat\beta}(\hat{s}_i-\hat{s}_j)$ that a held-out match between $i$ and~$j$ was won by~$i$, with $f_{\alpha\beta}(s)$ as in Eq.~\eqref{eq:fab}.  Further discussion of the procedure is given in Appendix~\ref{app:point-estimation}.

We can quantify the performance of our predictions by computing the log-likelihood of the actual outcomes of the held-out matches under the predicted probabilities~$\hat{p}_{ij}$.  If $W_{ij}$ is the number of times that $i$ actually won against~$j$ then the log-likelihood per game is
\begin{align}
Q = \frac{\sum_{ij} W_{ij} \log \hat{p}_{ij}}{\sum_{ij} W_{ij}}.
\label{eq:loglikelihoodpergame}
\end{align}
This measure naturally rewards cases where the model is confident in the correct answer ($\hat{p}_{ij}$~close to~1) and heavily penalizes cases where the model is confident in the wrong answer ($\hat{p}_{ij}$~close to~0).  Note that the log-likelihood is equal to minus the description length of the data---the amount of information needed to describe the true sequence of wins and losses in the held-out data given the estimated probabilities~$\hat{p}_{ij}$---so models with high log-likelihood are more parsimonious in describing the true pattern of wins and losses.

To place the performance of our proposed model in context, we compare it against a basket of other ranking models and methods, including widely used standards, some recently proposed approaches, and some variants of the approach proposed in this paper.  As a baseline we compare performance against the standard Bradley-Terry model with a logistic prior, which is commonly used in many ranking tasks, particularly in sports, and which we have ourselves used and recommended in the past~\cite{Newman23b}.  We measure the performance of all other models against this one by calculating the difference in the log-likelihood per match, Eq.~\eqref{eq:loglikelihoodpergame}.  The other models we test are:
\begin{enumerate*}
\item The luck-plus-depth model of this paper.
\item A depth-only variant in which the parameter $\alpha$ is set to zero.
\item A luck-only variant in which the parameter $\beta$ is set to $\infty$, which is equivalent to minimum violations ranking as described in Section~\ref{sec:combined}.
\item The Bradley-Terry model under maximum-likelihood estimation, which is equivalent to imposing an improper uniform prior.
\item The ``SpringRank'' model of De Bacco~\etal~\cite{DLM18}, which ranks competitors using a physically motivated mass-and-spring model.
\end{enumerate*}
The proportion of data held out in the cross-validation was 20\% in all cases, chosen uniformly at random, and at least 50 random repetitions of the complete process were performed for each model for each of the data sets listed in Table~\ref{tab:data}.

\begin{figure*}
\centering
\includegraphics[width=0.6\linewidth]{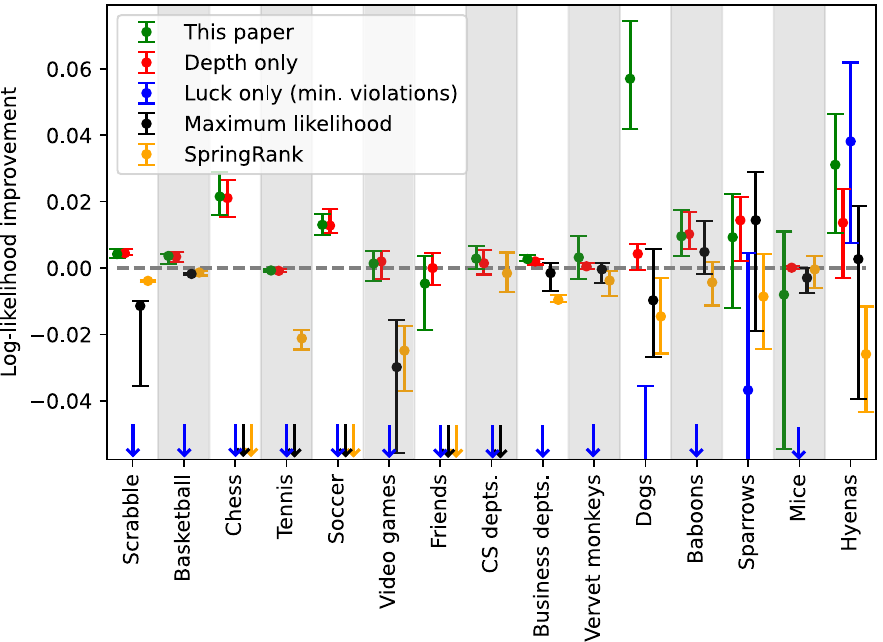}
\caption{Comparative performance of the model of this paper and a selection of competing models and methods, in the task of predicting the outcome of unobserved matches in a cross-validation experiment.  Performance is measured in terms of the log-likelihood (base~2) of the actual outcomes of matches within the fitted model, which is also equal to minus the description length in bits required to transmit the win/loss data given the fitted model.  Log-likelihoods are plotted relative to that of the standard Bradley-Terry model with a logistic prior (the horizontal dashed line).  Error bars represent upper and lower quartiles over at least 50 random repetitions of the cross-validation procedure in each case.  The arrows along the bottom of the plot indicate cases where the log-likelihood is outside the range of the plot.}
\label{fig:log-likelihood}
\end{figure*}

The results are summarized in Fig.~\ref{fig:log-likelihood}.  The horizontal dashed line in the figure represents the baseline set by the Bradley-Terry model and the points with error bars represent the increase (or decrease) in log-likelihood relative to this level for each model and data set.  The error bars represent the upper and lower quartiles of variation of the results over the random repetitions.  (We use quartiles rather than standard deviations because the distributions are highly non-normal in some cases.)

We note a number of things about these results.  First, the model of this paper performs best for every data set without exception, within the statistical uncertainty, although the depth-only version of the model is also competitive in many cases, particularly for the sports and games.  The latter observation is unsurprising, since, as we have said, there is little evidence for $\alpha>0$ in the games.  For the particular case of the dominance hierarchy of hyenas, the minimum violations ranking is competitive, which is also unsurprising: as shown in Fig.~\ref{fig:results} this hierarchy is very deep---the value of $\beta$ is over 100---and hence our model and the minimum violations ranking are essentially equivalent.  In all the other networks the minimum violations ranking performs worse---usually much worse---than our model.  (Arrows at the bottom of the figure indicate results so poor they fall off the bottom of the scale.)  The maximum likelihood fit to the Bradley-Terry model also performs quite poorly, a notable observation given that this is one of the most popular ranking algorithms in many settings.  It even performs markedly worse than the same Bradley-Terry model with a logistic prior.  Finally, we note that the SpringRank algorithm of~\cite{DLM18} is relatively competitive in these tests, though it still falls short of the model of this paper and the standard Bradley-Terry model with logistic prior.

\section{Conclusions}
In this paper we have studied the ranking of competitors based on pairwise comparisons between them, as happens for instance in sports, games, and social hierarchies.  Building on the standard Bradley-Terry ranking model, we have extended the model to include two additional features: an element of luck that allows weak competitors to occasionally beat strong ones, and a ``depth of competition'' parameter that captures the number of distinguishable levels of play in a hierarchy.  Deep hierarchies with many levels correspond to complex games or social structures.  We have fitted the proposed model to data sets representing social hierarchies among both humans and animals and a range of sports and games, including chess, basketball, soccer, and video games.  The fits give us estimates of the luck and depth of competition in each of these examples and we find a clear pattern in the results: sports and games tend to have shallow depth and little evidence of a luck component, while social hierarchies are significantly deeper and more often have an element of luck, with the animal hierarchies being deeper than the human ones.

We also test our model's ability to predict the outcome of contests.  Using a cross-validation approach we find that the model performs as well as or better than every other model tested in predictive tasks and very significantly better than the most common previous methods such as maximum likelihood fits to the Bradley-Terry model or minimum violations rankings.

\begin{acknowledgments}
The authors thank Elizabeth Bruch, Fred Feinberg, and Dan Larremore for useful conversations.  This work was funded in part by the US National Science Foundation under grant DMS--2005899.  All empirical data used in this paper are freely available online or from their original authors.
\end{acknowledgments}

\appendix
\section*{Appendices}

\subsection{Data sets}
\label{app:datasets}
The example data sets used in this paper are summarized in Table~\ref{tab:data} of the main paper and divide into three broad categories: sports and games (six data sets), human social hierarchies (three data sets), and animal social hierarchies (six data sets).  Here we provide some additional details on these data. 

\smallskip\textbf{Sports and games:} We consider both team competition (basketball, soccer) and individual competition (chess, Scrabble, tennis, video games).  For the team sports we treat each team in each year as a different entity with its own assigned score~$s_i$.  Thus, for example, the England soccer team in 2015 is considered a different entity from the England soccer team in 2014.  This reflects the fact that the composition of teams can change from season to season and with it the ranking of the team in comparison to others.

Two of the game data sets, for chess and Scrabble, were too large in their original form to perform our full Bayesian analysis in a reasonable amount of time, so they were subsampled to reduce them to manageable size.  We limited the chess data set to only those players who had participated in at least 200 games and then randomly selected 5\% of those players.  All others were removed from the data set. The scrabble data set was similarly pared down by limiting it to players who had at least 100 games and then choosing a random 20\% of those who remained.

Another issue with some of the game data is the presence of ties, which occur with moderate frequency in both chess and soccer.  Although there do exist ranking models that allow for ties~\cite{RK67,Davidson70}, we avoid these in the present work for the sake of simplicity, and all our models assume that the only possible outcomes of a match are a win or a loss.  To accommodate the chess and soccer data within this setting we remove all ties from the data, which amounts to 10--30\% of matches in those data sets. 

\textbf{Human social hierarchies:} A related issue arises in the ``friends'' data set, which details friend nominations among students in a US middle/high school.  A~significant fraction of such nominations are reciprocal---two individuals each nominate the other as a friend~\cite{HK88,BN13}.  Such reciprocated nominations have been treated as ties in some previous analyses~\cite{Newman23b}, but here again we simply remove them.  Only unreciprocated friendships are recorded as a win for the person who receives the nomination.

\textbf{Animal hierarchies:} Data on animal dominance hierarchies is copious: this has been an active field of research for at least sixty years.  The data sets studied in this paper come from a variety of sources, but particularly from DomArchive, a collection of 436 dominance interaction data sets compiled by Strauss~\etal~\cite{SDGHSC22}.  Data sets in the archive vary widely in size, but the sets we focus on are ones with a relatively large number of interactions per individual, which improves the statistics and helps reduce uncertainty on the fitted values of the model parameters.

\subsection{Cross-validation}
\label{app:cross-validation-measures}
In the cross-validation results reported in the main paper we quantify predictive performance of the various models by calculating the log-likelihood of the testing (held-out) data within the fitted model---see Fig.~\ref{fig:log-likelihood}.  This is not, however, the only way to measure performance; there are a number of other approaches in common use.  In this appendix we describe some alternative performance metrics and investigate how our models size up when measured by these metrics.  In general the results are similar to those presented in the main paper, but there are some differences in the details. 

A simple way to quantify the predictive performance of a model is to count the number of times the model predicts the correct winner in the test data.  As before, we start by fitting the model to the training portion of the data to obtain MAP estimates $\hat{\vec{s}}$ of the scores, then, given those estimates, player~$i$ is considered favored to beat player~$j$ if $\hat{s}_i > \hat{s}_j$.  The \textit{accuracy}~$C$ of the model is defined to be the fraction of matches in the testing data where this prediction is born out:
\begin{align}
    C = \frac{\sum_{ij} W_{ij} \mathbf{1}_{\hat{s}_i > \hat{s}_j}}{\sum_{ij} W_{ij}}
\end{align} 
where $W_{ij}$ is the number of times $i$ beats $j$ in the testing data, as previously, and $\mathbf{1}_x$ is the indicator function which is 1 if $x$ is true and 0 otherwise. 

\begin{figure}
\centering
\includegraphics[width=\columnwidth]{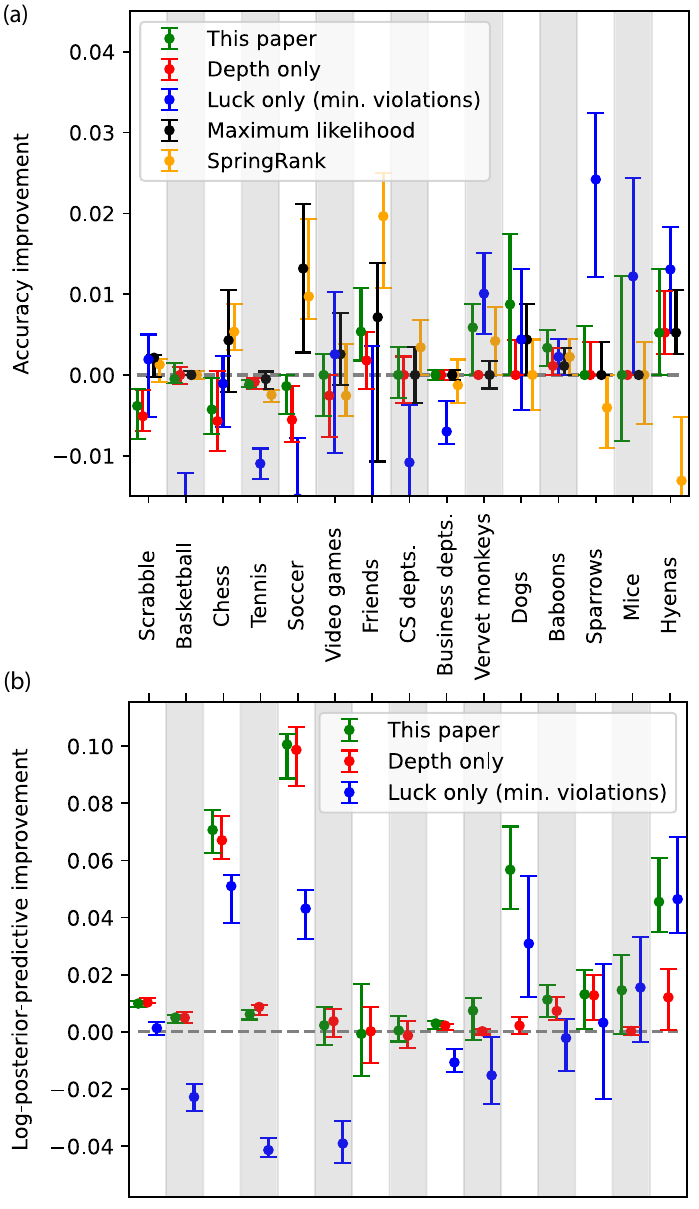}
\caption{Results from the same set of cross-validation tests shown in Fig.~\ref{fig:log-likelihood}, but quantified using (a)~accuracy and (b)~log-posterior predictive probability, instead of log-likelihood.  All results are measured relative to the Bradley-Terry model with a logistic prior, which is represented as the dashed horizontal line in each panel.  Error bars represent upper and lower quartiles, estimated from at least 50 random repetitions of the cross-validation procedure in each case.  The maximum likelihood and SpringRank models are not included in the lower comparison, since they are based on point estimates rather than Bayesian methods and hence one cannot calculate a posterior-predictive probability.}
\label{fig:accuracy-posterior-predictive}
\end{figure}

Values of this accuracy measure are shown in Fig.~\ref{fig:accuracy-posterior-predictive}a for each of the models considered in this paper for each of our data sets.  As with our previous results for log-likelihood, we report performance relative to a baseline set by the standard Bradley-Terry model with a logistic prior, represented by the horizontal dashed line in the figure.  Comparing with our earlier results from Fig.~\ref{fig:log-likelihood}, the difference between models is smaller when measured in terms of accuracy than log-likelihood.  For example, the minimum violations ranking performs quite poorly according to the log-likelihood, but is comparable and sometimes better than our models in terms of accuracy.  This may be because the minimum violations ranking is more directly tuned to solving this specific problem: by minimizing violations we precisely minimize the number of outcomes that are predicted incorrectly.  On the other hand, the minimum violations algorithm does not reflect how confident we are in each outcome or any other aspect of the prediction task, and in this sense is inferior to other approaches.

Both the likelihood and accuracy measures are based on point estimates of model parameters $\hat{\vec{s}}, \hat{\alpha}$, and $\hat{\beta}$ but, as shown in Fig.~\ref{fig:results}, point estimates do not always do a good job of capturing the full posterior distribution $P(\vec{s},\alpha,\beta|\Atrain)$, particularly in sparse data sets.  To get around this issue, we can calculate the average of the likelihood over the distribution of parameter values thus:
\begin{align}
P(&\Atest|\Atrain) \nonumber\\
    &= \int P(\Atest|\vec{s}, \alpha, \beta)P(\vec{s}, \alpha, \beta|\Atrain) \>d^n\!\vec{s} \>d\alpha \>d\beta.
\end{align} 
In practice, this quantity can be estimated from a set of $N$ samples of $(\vec{s}_k, \alpha_k, \beta_k)$ (with $k=1\ldots N$) drawn from the posterior $P(\vec{s}, \alpha, \beta|\Atrain)$, as the average
\begin{align}
    P(\Atest|\Atrain) \simeq \frac{1}{N} \sum_{k=1}^N P(\Atest|\vec{s}_k, \alpha_k, \beta_k).
\end{align}
We can calculate this estimate from the same Monte Carlo samples we already generated, which we used previously to visualize the posterior distribution in Fig.~\ref{fig:results}.  As our measure of performance we then compute the \textit{log-posterior-predictive probability} per game
\begin{align}
    R = \frac{\log P(\Atest|\Atrain)}{\sum_{ij} W_{ij}},
\end{align}
a fully Bayesian performance measure.

We plot this measure for a number of our models and data sets in Fig.~\ref{fig:accuracy-posterior-predictive}b.  Note, however, that since the measure involves an integral over the posterior distribution of the scores, we cannot apply it to ranking methods that return only point estimates of the scores rather than a full probability distribution, which in this case means the Bradley-Terry MLE and SpringRank, which are thus excluded from the figure.  Among the remaining methods the full luck-plus-depth model of this paper performs best, or equal-best, for every data set, by this measure.

\subsection{Point estimates of parameters}
\label{app:point-estimation}
To compute the log-likelihood and accuracy measures of predictive success we use point estimates of the model parameters and scores, which we compute one after the other: we estimate the expected posterior values of the parameters $\hat{\alpha}$, $\hat{\beta}$ from a simple average of the Monte Carlo samples, then we fix these values and compute the MAP values of the scores~$\hat{\vec{s}}$ using a standard numerical optimization method.  We could, alternatively, use the expected values of the scores, which would be easy to calculate from the samples, but we prefer MAP values since they give a more appropriate point of comparison with other approaches based on maximum probability estimates, such as the maximum likelihood fit to the Bradley-Terry model or the SpringRank algorithm.

One might imagine one could simplify the calculation by just jointly optimizing the posterior $P(\vec{s},\alpha,\beta|\mat{A})$ over both the scores and parameters to define estimates
\begin{align}
    (\vec{s}^*,\alpha^*,\beta^*) \equiv \text{argmax}_{\vec{s},\alpha,\beta} P(\vec{s},\alpha,\beta|\mat{A}).
\label{eq:alpha-star-beta-star}
\end{align}
We find, however, that this can give biased results by artificially inflating the value of the depth parameter~$\beta$.  This happens because the likelihood $P(\mat{A}|\vec{s},\alpha,\beta)$ is a function of the product $\beta \vec{s}$ (see Eq.~\eqref{eq:fab}), meaning that the value of the likelihood is unchanged if we increase $\beta$ while simultaneously reducing all the scores by the same factor.  Reducing $\vec{s}$ in this way increases the prior $P(\vec{s})$ (which is peaked at $\vec{s} = 0$) and so increases the posterior $P(\vec{s},\alpha,\beta|\mat{A})$.  Unchecked, this effect would send the joint maximum to $\beta^* \to \infty$, $\vec{s} \to 0$.  The prior $P(\beta)$ somewhat mitigates this problem, but in practice the jointly fitted value~$\beta^*$ is still unreasonably large: values for each of the data sets are shown in Table~\ref{tab:parameters}.

\subsection{Other measures of depth}
In this paper we measure depth of competition by the parameter~$\beta$ in our joint luck-plus-depth model, Eq.~\eqref{eq:fab}.  This is not the only possible approach for quantifying depth, however, and in this appendix we discuss some alternative approaches and explain how they relate to similar ideas presented elsewhere.

As discussed in Section~\ref{sec:depth}, our depth measure~$\beta$ counts the number of ``levels of skill'' between two typical players in a population, who in expectation have a priori score difference $s_i-s_j=1$ (because of our choice of prior on~$s$).  An alternative, and common, way to define depth is as the number of levels between not the typical pair of players but the best and worst players, which is given by
\begin{align}
    \hat{\beta}_{\text{range}} = \hat{\beta}(\hat{s}_{\text{max}} - \hat{s}_{\text{min}}).
\label{eq:beta-range}
\end{align}

In the data sets studied here we find that the factor $\hat{s}_{\text{max}} -\hat{s}_{\text{min}}$ varies from about 2.5 to~4.  The range tends to be larger when there are more competitors, presumably because outliers are more likely in large samples, and we regard this as downside of this measure, although in practice the depth order of our data sets does not change significantly between this measure and our own.  Values of $\hat{\beta}_{\text{range}}$ are reported in Table~\ref{tab:parameters} for each of the data sets. 

Our depth measure~$\beta$ is defined in the context of our full luck-plus-depth model, but in many cases, particularly for the sports data sets, there is no strong evidence of a nonzero luck parameter~$\alpha$.  An alternative approach for quantifying depth in these cases is to use a depth-only model as in Eq.~\eqref{eq:fbeta}.  Depth values calculated by fitting this model are given in Table~\ref{tab:parameters} and denoted $\beta_0$, which we refer to as ``restricted depth.''  In practice these figures are not very different for those for~$\beta$ in cases (such as sports) where the value of $\alpha$ is small anyway, or more precisely when the posterior distribution in Figure~\ref{fig:results} meaningfully intersects the $\alpha = 0$ axis, so that the zero-luck model is plausible.  On the other hand, $\beta$~and $\beta_0$ can differ substantially when the data support a significantly nonzero value of~$\alpha$.  For example, the mice data set has an expected value of $\alpha$ around 0.25 with a posterior distribution that has considerable separation from $\alpha = 0$, and in this case we find a large difference between a value of $\hat\beta=26.5$ and $\hat{\beta}_0 = 2.1$, the latter being more akin to the sports data than to the other animal hierarchies.

The restricted depth~$\beta_0$ is closer in spirit to previous measures of depth that do not consider the element of luck, and the occurrence of large discrepancies with the value of~$\beta$ in some data sets suggests that such previous measures might potentially be in error by a significant margin.  For applications where the element of luck is not an issue, however, the restricted depth could be useful as a simplification of our measure.  It can be calculated relatively straightforwardly, to a good approximation, using the standard Bradley-Terry model with a logistic prior, a model we have recommended in the past.  In our current analysis we have used Gaussian priors, but the logistic prior has some practical advantages in that it enables simple and fast iterative methods for computing MAP scores.  In the most common version of this approach, one uses the unit logistic distribution $1/[(1+e^s)(1+e^{-s})]$ as prior with the standard ($\beta=1$) Bradley-Terry model, which leads to an elegant iterative algorithm for calculating the scores~\cite{Newman23b}.  The logistic prior, however, has variance $\frac13 \pi^2$, whereas our Gaussian prior has variance~$\frac12$, so, though the qualitative shape of the two distributions is similar, the logistic distribution has substantially greater width, by a factor of $\pi\sqrt{2/3}$.  An alternative way to perform the same calculation is to shrink the width of the prior to be the same as the Gaussian, while simultaneously shrinking the width of the Bradley-Terry score function by the same factor, which is equivalent to choosing $\beta = \pi\sqrt{2/3} = 2.565$.  This leaves the algorithm, and the resulting ranking, unchanged, and thus the iterative method with a logistic prior is equivalent to the depth-only model with $\beta = 2.565$.

Happily, this choice of $\beta$ falls squarely in the middle of the range of values seen in Fig.~\ref{fig:results} and in practice this approach has quite competitive performance, as shown in Fig.~\ref{fig:log-likelihood}, where it is used as the baseline.  On the other hand, there are plenty of cases where the value $\beta = 2.565$ is clearly misspecified, which is signaled by fitted scores whose variance does not match the width of the prior.  This observation suggests that we could use the spread of the fitted scores as a heuristic measure of (restricted) depth and in practice this approach seems to work quite well.  Quantifying the spread by its the standard deviation, we report figures for each of our data sets in Table~\ref{tab:parameters}, and we find that there is good correlation between this standard deviation and the restricted depth~$\hat{\beta}_0$ as calculated earlier.  Given that the former is significantly easier to calculate than the latter, this could be a useful approach for calculations where accuracy and rigor are not at a premium.

A quite different approach to measuring depth has been developed in the animal behavior literature, where the notion of ``steepness'' has gained currency in discussions of dominance hierarchies~\cite{DSV06}. Steepness is most often defined through quantities known as ``David's scores,'' which are measures of individual performance analogous to our fitted $s_i$~\cite{David88}.  The David's scores are defined as
\begin{align}
    \text{DS}_i = w_i + \sum_j w_j P_{ij} - l_i - \sum_j l_j P_{ji}
\end{align}
where $P_{ij}$ is the fraction of times that $i$ beats~$j$:
\begin{align}
    P_{ij} = \frac{A_{ij}}{A_{ij}+A_{ji}},
\end{align}
and $w_i$ and $l_i$ are row and column sums of this matrix:
\begin{align}
    w_i = \sum_j P_{ij}, \qquad l_i = \sum_j P_{ji}. 
\end{align}
De Vries~\etal~\cite{DSV06} propose normalizing the David's scores according to
\begin{align}
    \text{NormDS}_i = {\text{DS}_i + {n\choose2}\over n},
\end{align}
which vary between 0 and $n-1$, then the animals are ranked according to the resulting values.  With the inferred rank order on the $x$-axis and the normalized David's score on the $y$-axis, the steepness of the hierarchy is then defined to be the slope~$S_{\text{DS}}$ of the ordinary line of best fit.  A nice feature of this formulation is that the steepness runs from 0 to 1, with the value 1 being achieved in any hierarchy where all dominance interactions run from higher ranked to lower ranked individuals (zero violations).

Neumann and Fischer~\cite{NF23} have recently proposed a related measure that considers the slope~$S_\text{Elo}$ of the line of best fit between Elo scores for the competitors and their inferred ordinal ranking.  Elo scores are essentially a sequential (time-dependent) version of a maximum likelihood fit to the Bradley-Terry model and so this definition is closer to the ideas considered in this paper.  Neumann and Fischer also incorporate Bayesian elements where certain aspects of the fitting process are randomized, such as the sequential order (if the true order is unknown) and the initial values of the ratings.

In Table~\ref{tab:parameters} we report values for a number of our data sets of $S_{\text{DS}}$ (calculated using the R package \verb|steepness|~\cite{LV22}) and $S_{\text{Elo}}$ (calculated using the R package \verb|EloSteepness|~\cite{Neumann23}).  Overall, we find that the results are clearly correlated with the other measures shown in the table, although $S_{\text{DS}}$ has trouble differentiating between the lower depth data sets.  The Elo-based steepness $S_{\text{Elo}}$ fares better and correlates quite well with the restricted depth~$\hat{\beta}_0$, although the calculations are computationally demanding on account of the randomization and prove intractable for our larger data sets (as indicated by ``--'' in the table). 

To complete our collection of measures of depth we also include in Table~\ref{tab:parameters} the parameter~$\beta_S$ that appears in the SpringRank model~\cite{DLM18}.  This parameter has not previously been used as a measure of depth but one can make an argument for its use in this way---see Appendix~\ref{app:springrank}.

Finally, we note in passing that there is an analogy between the depth parameter~$\beta$ and a notion of ``temperature'' for a data set.  The form of the score function of Eq.~\eqref{eq:fbeta} is precisely that of the Fermi-Dirac probability function of many-body physics, the probability of occupation at inverse temperature~$\beta$ of an energy level with energy~$s$ above the Fermi level.  While we have not directly exploited this analogy here, it is a part of a broader correspondence between noise and unpredictability in statistics and temperature in physics.

\begin{table*}[htbp]
\centering
    \begin{tabular}{l|l|rrrrrrrr|rr}
        & & & \multicolumn{6}{c}{Measures of depth} & & \multicolumn{2}{c}{Luck} \\
        & Data set & $\hat{\beta}$ & $\beta^*$ & $\hat{\beta}_{\text{range}}$ & $\hat{\beta}_0$ & $\text{std}(\hat{\vec{s}}_L)$ & $S_{\text{DS}}$ & $S_{\text{Elo}}$  & $\hat{\beta}_S$ & $\hat{\alpha}$ & $\alpha^*$ \\
        \hline
        \hline
        \multirow{7}{*}{\rotatebox[origin=c]{90}{\hspace{10pt}Sports/games}} & Scrabble & 0.68 & 3.13 & 2.43 & 0.60 & 0.64 &  0.00 & -- & 2.24 &0.09 & 0.00 \\
        & Basketball & 1.01 & 10.79 & 3.66 & 0.83 & 0.61 & 0.01 & 0.48 & 2.32 & 0.13 & 0.02 \\
        & Chess & 1.17 & 4.73 & 4.21 & 1.04 & 0.91 & 0.00 & -- & 2.85 & 0.07 & 0.12 \\
        & Tennis & 1.44 & 1.98 & 5.88 & 1.34 & 0.72 & 0.00 & -- & 2.67 & 0.04 & 0.00 \\
        & Soccer & 1.73 & 6.23 & 4.97 & 1.58 & 1.02 & 0.00 & -- & 4.00 & 0.04 & 0.00 \\
        & Video games & 1.77 & 17.53 & 5.12 & 1.55 & 1.10 & 0.02 & 0.62 & 2.95 & 0.07 & 0.05 \\
        \hline
        \multirow{3}{*}{\rotatebox[origin=c]{90}{Human}} & Friends & 3.54 & 10.36 & 9.88 & 2.80 & 1.16 & 0.00 & -- & 5.23 & 0.05 & 0.00 \\
        & CS departments & 4.25 & 15.42 & 12.11 & 3.88 & 1.88 & 0.01 & 0.78 & 4.46 & 0.01 & 0.00 \\
        & Business depts. & 4.36 & 13.72 & 11.73 & 4.07 & 2.25 & 0.14 & 0.84 & 4.07 & 0.01 & 0.01 \\
        \hline
        \multirow{6}{*}{\rotatebox[origin=c]{90}{Animal}} & Vervet monkeys & 6.01 & 30.39 & 17.07 & 3.57 & 2.23 & 0.40 & 0.85 & 4.34 & 0.07 & 0.07 \\
        & Dogs & 8.74 & 33.29 & 24.82 & 3.76 & 2.03 & 0.25 & 0.93 & 3.65 & 0.11 & 0.09 \\
        & Baboons & 13.19 & 18.61 & 39.04 & 9.37 & 4.38 & 0.05 & 0.95 & 5.63 & 0.02 & 0.02 \\
        & Sparrows & 22.92 & 63.89 & 69.68 & 8.68 & 3.62 & 0.50 & 0.91 & 7.72 & 0.02 & 0.01 \\
        & Mice & 26.48 & 59.48 & 72.29 & 2.10 & 1.35 & 0.31 & 0.72 & 3.22 & 0.25 & 0.24 \\
        & Hyenas & 100.58 & 168.48 & 246.42 & 9.83 & 4.00 & 0.30 & 0.95 & 8.15 & 0.02 & 0.02 
    \end{tabular}
\caption{Inferred parameter values for the data sets considered in Section~\ref{sec:results}.  From left to right: $\hat{\beta}$ is expected depth, $\beta^*$ is the jointly optimized MAP depth as in Eq.~\eqref{eq:alpha-star-beta-star}, $\hat{\beta}_{\text{range}}$ is depth between the best and worst player as in Eq.~\eqref{eq:beta-range}, $\hat{\beta}_0$ is restricted depth as inferred in the depth-only $(\alpha = 0)$ model, $\text{std}(\hat{\vec{s}}_L)$ is the standard deviation of the MAP scores within the logistic-prior model, $S_{\text{DS}}$ is the steepness measure of de Vries~\etal~\cite{DSV06}, $S_{\text{Elo}}$ is the steepness measure of Neumann and Fischer~\cite{NF23}, $\hat{\beta}_S$~is the maximum likelihood estimate of the parameter~$\beta_S$ in the SpringRank model~\cite{DLM18}, $\hat{\alpha}$ is the expected luck, and $\alpha^*$ is the jointly optimized MAP estimate of the luck.}
    \label{tab:parameters}
\end{table*}

\subsection{Depth as predictability}
\label{app:depth-predictability}
In Section~\ref{sec:results} we observed that among our data sets the sports and games have lower depth compared to the social hierarchies, and we speculated that this was because a high-depth sport would not be as interesting to watch: at high depth a typical pair of competitors will be very unevenly matched and there will be little suspense about who is going to win.  In other words, high depth should result in high predictability of outcomes.  In this appendix we test this hypothesis by calculating various measures of predictability.

A natural measure of predictability is the same log-likelihood that we studied in Section~\ref{sec:outcomes}.  The log-likelihood of a data set is equal to minus the description length of the outcomes of the matches in that set, given the fitted model.  That is, it is equal to the amount of information it would take to communicate the outcomes to a receiver who already knows the fitted model.  Higher information (more negative log-likelihood) implies more unpredictable outcomes.  Completely random outcomes (matches decided by the toss of a coin) would give a log-likelihood of $-1$ per match (in log-base-2 units), while completely predictable ones would give zero.

\begin{figure}
\centering
\includegraphics[width=\linewidth]{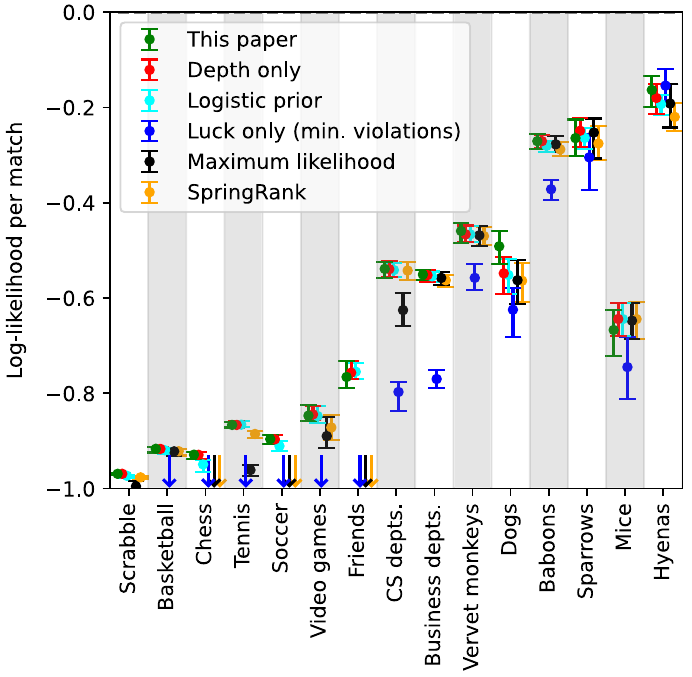}
\caption{Absolute log-likelihood values per match in the cross-validation tests of Fig.~\ref{fig:log-likelihood}.  This figure differs from Fig.~\ref{fig:log-likelihood} in showing absolute values rather than values relative to the Bradley-Terry model with logistic prior.}
\label{fig:log-likelihood-abs}
\end{figure}

Previously, we plotted the log-likelihood relative to the baseline set by the standard Bradley-Terry model, but in the present context we are interested in the absolute value.  Figure~\ref{fig:log-likelihood-abs} shows the absolute value for each of our data sets, arranged in order of increasing depth~$\beta$.  As the figure shows, the low-depth sports on the left are indeed quite unpredictable and none of our models perform much better than chance at predicting outcomes (log-likelihood per match is close to~$-1$).  As depth increases, however, outcomes generally become more predictable, and the deepest animal hierarchies have a log-likelihood approaching zero, meaning outcomes are nearly perfectly predictable.

There are some exceptions to this trend, most notably the mice data set which, as seen in Fig.~\ref{fig:results}, has a large element of luck $(\hat{\alpha} \simeq 0.25)$.  This introduces substantial randomness into the matches, despite the high depth, and greatly decreases predictability.

\begin{figure}
\centering
\includegraphics[width=\columnwidth]{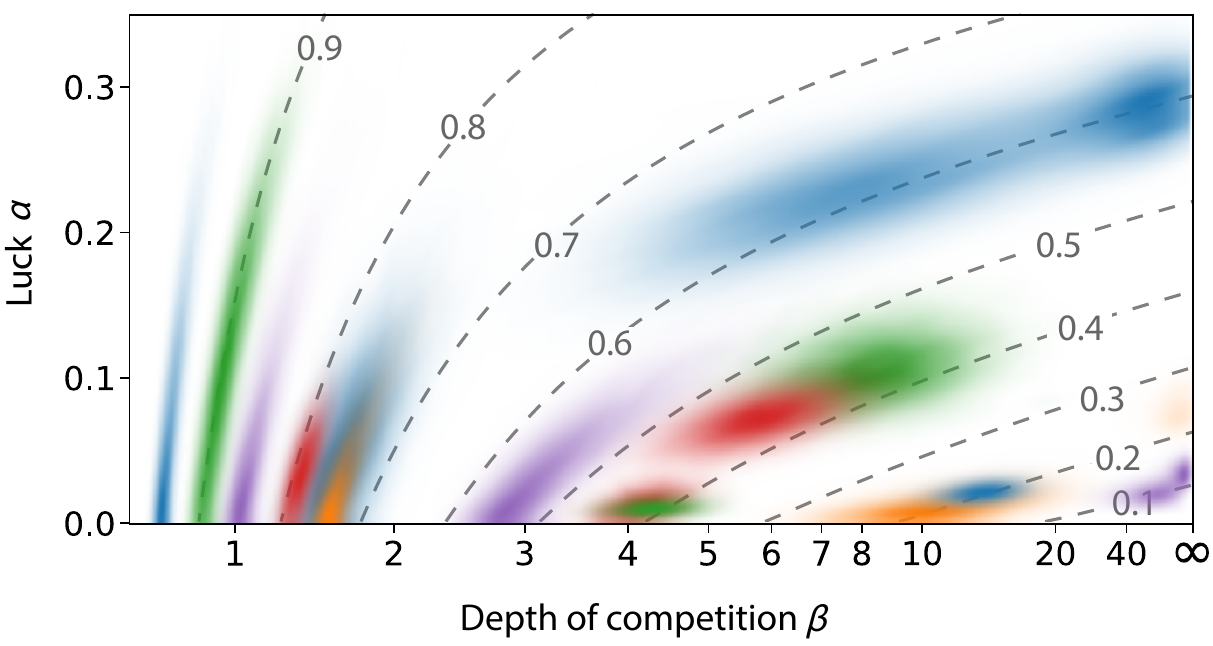}
\caption{The data sets of Fig.~\ref{fig:results} with dashed lines representing the contours of average entropy per match.  Low entropy indicates confidence about the outcome of a match; high entropy indicates unpredictability.}
\label{fig:entropy-alpha-beta}
\end{figure}

We can shed further light on predictability by calculating the average amount of information needed to describe matches that are truly drawn from our model.  That is, we consider two players whose scores~$s_i$ are drawn from our normal prior with variance~$\frac12$, so that the difference of their scores is normally distributed with variance~1, and we assume that the probability of $i$ beating $j$ is given exactly by $p_{ij} = f_{\alpha\beta}(s_i-s_j)$, Eq.~\eqref{eq:fab}, for some values of $\alpha$ and $\beta$ that we specify.  Then the average information needed to describe the outcome of the match is given by the standard entropy function for a Bernoulli random variable
\begin{align}
H[p_{ij}] = -p_{ij} \log p_{ij} - (1-p_{ij})\log(1-p_{ij}).
\end{align}
Then, writing $s = s_i-s_j$ and integrating, the average entropy per match over matches between many random pairs of players is
\begin{align}
S_{\alpha\beta} = {1\over\sqrt{2\pi}} \int_{-\infty}^\infty H\bigl[f_{\alpha\beta}(s)\bigr]\,e^{-s^2/2} \>ds.
\label{eq:entropyint}
\end{align}
Unfortunately, this integral does not seem to have a closed-form solution, but it can be evaluated numerically.  Figure~\ref{fig:entropy-alpha-beta} shows a modified version of Fig.~\ref{fig:results} from the main paper, representing the posterior probability distribution of $\alpha,\beta$ for our various data sets, with superimposed lines representing the contours of the average entropy.  As the figure shows, the entropy is higher for lower depth and for higher luck, as we would expect, since both increase the unpredictability of outcomes.  We also note that the posterior distributions of individual data sets appear to follow the contour lines quite closely, arcing upward and to the right.  This occurs because the entropy is by definition equal to minus the log-likelihood, and our prior on $\alpha$ and $\beta$ is slowly varying by construction, so the posterior is also slowly varying along the contour lines of constant likelihood.  The contour lines are calculated as averages over outcomes drawn from the fitted model, whereas the probability clouds in the figure represent real-world data, so the two are not precisely comparable.  But to the extent that the data are well described by the model we would expect them to agree and hence for the clouds to follow the contours in the plot.  This also means that, while some of the clouds in the figure are quite extended, indicating substantial uncertainty about the values of $\alpha$ and~$\beta$, they are narrow in the direction perpendicular to the contours, meaning that we have high confidence about the value of the log-likelihood.  This is reflected in Fig.~\ref{fig:log-likelihood-abs}, where we see that the uncertainty on our estimates of the log-likelihood is quite modest.

\subsection{SpringRank}
\label{app:springrank}
Among the various approaches to ranking considered in this paper, SpringRank~\cite{DLM18} is a recent and novel approach based on a physical analogy to the behavior of a network of masses and springs.  In this appendix we make some observations on the method and how it relates to the Bradley-Terry model, which forms the foundation for the other methods we consider.

In SpringRank the likelihood of observing a directed network~$\mat{A}$ is given by a product of Poisson distributions over all possible directed edges:
\begin{align}
P(&\mat{A}|\vec{s},\beta_S,c) = \prod_{ij} \frac{ r_{ij}^{A_{ij}}}{A_{ij}!} e^{-r_{ij}},
\label{eq:PAsbc}
\end{align}
with the expect number of directed edges~$i\to j$ given by
\begin{equation}
    r_{ij} = ce^{-\frac12\beta_S(s_i-s_j-1)^2},
\end{equation}
for given scores~$\vec{s}$, inverse temperature~$\beta_S$, and a ``sparsity'' parameter~$c$. 
Equation~\eqref{eq:PAsbc} can be rewritten as
\begin{align}
P&(\mat{A}|\vec{s},\beta_S,c)
  = \prod_{i<j} \frac{r_{ij}^{A_{ij}}}{A_{ij}!} e^{-r_{ij}}\>
  \frac{r_{ji}^{A_{ji}}}{A_{ji}!} e^{-r_{ji}} \nonumber\\
  &= \prod_{i<j} \frac{(r_{ij}+r_{ji})^{A_{ij} + A_{ji}}e^{-(r_{ij}+r_{ji})}}{(A_{ij} + A_{ji})!} \nonumber\\
  &\qquad{}\times\frac{(A_{ij} + A_{ji})!}{A_{ij}!A_{ji}!}\left(\frac{r_{ij}}{r_{ij}+r_{ji}}\right)^{A_{ij}}\left( \frac{r_{ji}}{r_{ij}+r_{ji}}\right)^{A_{ji}} \nonumber\\
  &= \prod_{i<j} \frac{ m_{ij}^{\bar{A}_{ij}}e^{-m_{ij}}}{\bar{A}_{ij}!}  \nonumber\\
  &\quad{}\times\binom{\bar{A}_{ij}}{A_{ij}} \frac{1}{[1 + e^{-2\beta_S(s_i - s_j)}]^{A_{ij}}[1 + e^{-2\beta_S(s_j - s_i)}]^{A_{ji}}},
\label{eq:PAsbc3}
\end{align}
where $m_{ij} = r_{ij}+r_{ji}$ and $\bar{A}_{ij} = A_{ij}+A_{ji}$ is an element of the adjacency matrix~$\bar{\mat{A}}$ of the undirected network of matches.

Equation~\eqref{eq:PAsbc3} is equal to the likelihood of generating an undirected network~$\bar{\mat{A}}$ of matches and then separately choosing the directions of the edges, i.e.,~the winners of the matches:
\begin{equation}
P(\mat{A}|\vec{s},\beta_S,c)
    = P(\bar{\mat{A}}|\vec{s},\beta_S,c)\,
      P(\mat{A}|\vec{s},\beta_S,\bar{\mat{A}}),
\label{eq:SRsplit}
\end{equation}
where the probability of the undirected network is another product of Poisson distributions:
\begin{equation}
P(\bar{\mat{A}}|\vec{s},\beta_S,c)
  = \prod_{i<j} \frac{ m_{ij}^{\bar{A}_{ij}}e^{-m_{ij}}}{\bar{A}_{ij}!}
\label{eq:PAsbc2}
\end{equation}
and
\begin{align}
P&(\mat{A}|\vec{s},\beta_S,\bar{\mat{A}}) \nonumber\\
  &= \prod_{i<j} \binom{\bar{A}_{ij}}{A_{ij}} \frac{1}{[1 + e^{-2\beta_S(s_i - s_j)}]^{A_{ij}}[1 + e^{-2\beta_S(s_j - s_i)}]^{A_{ji}}}.
\label{eq:PAsbA}
\end{align}
(It is straightforward to confirm that the latter is correctly normalized for $A_{ij} = 0\ldots\bar{A}_{ij}$ and $A_{ji} = \bar{A}_{ij}-A_{ij}$.)

But Eq.~\eqref{eq:PAsbA} is identical to the likelihood for the model studied in this paper, Eqs.~\eqref{eq:PAGsf} and~\eqref{eq:fab}, with $\alpha=0$ and $\beta=2\beta_S$.  (The binomial coefficient accounts for the number of ways of assigning directions~$A_{ij}$ to the $\bar{A}_{ij}$ undirected edges.)  This observation suggests that we might use $\beta_S$ as a a measure of the (restricted) depth of a hierarchy, and indeed we observe a correlation between the maximum likelihood value~$\hat{\beta}_S$ and our own restricted depth parameter~$\beta_0$, as shown in Table~\ref{tab:parameters}. 

\begin{figure}
\centering
\includegraphics[width=0.9\columnwidth]{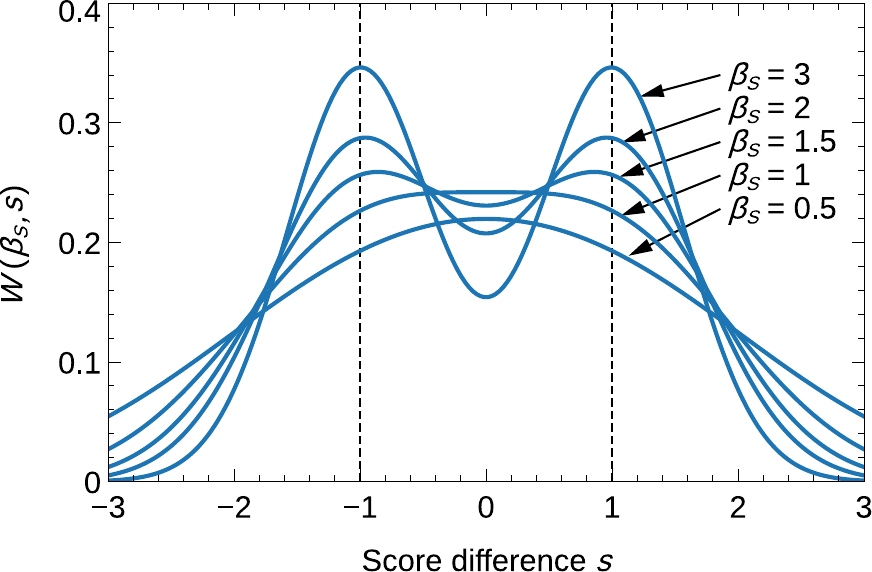}
\caption{The function $W(\beta_S,s)$ of Eq.~\eqref{eq:defsW} plotted against~$s$, for various values of $\beta_S$ as indicated.}
\label{fig:Wbetas}
\end{figure}

However, it is the other term, Eq.~\eqref{eq:PAsbc2}, that particularly distinguishes SpringRank from the other models we have considered.  This term, which measures the likelihood that the set of observed matches occurs at all, has no equivalent in the Bradley-Terry model and related models.  The quantity $m_{ij}$, which is the expected number of matches between $i$ and $j$, can be rewritten in the form
\begin{align}
m_{ij} = M \frac{W(\beta_S, s_i - s_j)}{\sum_{i < j} W(\beta_S, s_i - s_j)},
\label{eq:mij}
\end{align}
where
\begin{align}
W(\beta_S,s) = \sqrt{\frac{\beta_S}{8 \pi}} \bigl[e^{-\frac12 \beta_S (s - 1)^2} + e^{-\frac12 \beta_S (s + 1)^2}\bigr].
\label{eq:defsW}
\end{align}
(Note that $W(\beta_S,s)$ is symmetric in~$s$ so the sign of the score difference in Eq.~\eqref{eq:mij} has no effect.)  In this formulation the parameter $M$ controls the total number of (undirected) edges in the network and the (properly normalized) probability density $W(\beta_S,s_i - s_j)$ controls how they are distributed given the scores~$s_i$.  Figure~\ref{fig:Wbetas} shows the form of $W(\beta_S, s)$ for various choices of~$\beta_S$.  For $\beta_S\le1$ there is a single peak at $s=0$ so that interactions are preferentially between evenly matched players, but above $\beta_S=1$ the function becomes bimodal and increasingly peaked around $s = \pm1$, so that players with a score difference near~1 are more likely to interact.

It is arguably a disadvantage of the SpringRank model that the same parameter~$\beta_S$ controls both the depth of competition via Eq.~\eqref{eq:PAsbA} and the distribution of matches via Eq.~\eqref{eq:mij}.  Conceptually these are separate processes, and one could make an argument for a model in which they were controlled by separate parameters, although we have not taken that approach here---we use the model as originally defined for the sake of consistency.

In our cross-validation tests we use the maximum likelihood point estimate for the value of $\beta_S$, in keeping with the other models we study.  We note, however, that De~Bacco \etal~\cite{DLM18}, in their original work on SpringRank, used different values of $\beta_S$ depending on whether the results were scored using log-likelihood or accuracy, choosing in each case the value that gave the best performance according to the measure used.

Finally, we note that the original specification of the SpringRank model also included an optional Gaussian prior on the scores.  We have not adopted this prior in our tests, since we find that it tends to diminish the performance of the method.

\end{document}